\documentclass[10pt]{article}

\begin{document}

\title{Effective spacetime from multi-dimensional gravity}
\author{J. Ponce de Leon\thanks{E-Mail:
jpdel@ltp.upr.clu.edu, jpdel1@hotmail.com}  \\Laboratory of Theoretical Physics, 
Department of Physics\\ 
University of Puerto Rico, P.O. Box 23343,  
San Juan,\\ PR 00931, USA}
\date{November 2009}

\maketitle
\begin{abstract}
We study the effective spacetimes  in lower dimensions  that can be extracted from a multidimensional generalization of the Schwarzschild-Tangherlini spacetimes derived by Fadeev, Ivashchuk and Melnikov ({\it Phys. Lett,} {\bf A 161} (1991) 98). The higher-dimensional spacetime has $D = (4 + n + m)$ dimensions, where $n$ 
and $m$ are the number of ``internal" and ``external" extra dimensions, respectively. We analyze the effective $(4 + n)$ spacetime obtained after dimensional reduction of the $m$ external dimensions.   We find that when the $m$ extra dimensions are compact (i) the physics in lower dimensions   is independent of $m$ and the character of the singularities in higher dimensions, and (ii) the total gravitational mass $M$ of the effective matter distribution  is  less than the Schwarzshild mass.  
In contrast, when the $m$  extra dimensions are large this is not so; the physics in $(4 + n)$ does explicitly depend on $m$, as well as on the  nature of the singularities in high dimensions,  and the mass of the effective matter distribution (with the exception of wormhole-like distributions) is bigger than the Schwarzshild mass.  These results may be relevant to  observations for an experimental/observational test of the theory.

\end{abstract}

\medskip



\newpage

\section{Introduction}
Nowadays, there are a number of theories suggesting that the universe may have more than  four dimensions. These arise naturally in supergravity $(11D)$ and superstring theories $(10D)$, which seek the unification of gravity with the interactions of particle physics, and are expected to become important near the horizon of black holes, as ``windows" to extra dimensions \cite{Davidson Owen}, and during the  evolution of the early universe \cite{evol of early universe}. Also, it appears that black holes will play a crucial role in understanding non-perturbative effects in a quantum theory of gravity \cite{Myers}. The natural question here is  to know  which of the properties of black holes are particular only to $(3 + 1)$ dimensions,  and which hold more generally. 

Higher dimensional extensions of the Schwarzschild black hole metric  have been obtained by Tangherlini \cite{Tangherlini} and generalized by Myers and Perry \cite{Myers}.    Fadeev, Ivashchuk and Melnikov \cite{Fadeev} obtained a class of static, spherically symmetric solutions of the Einstein vacuum field equations, which generalize the Tangherlini solution to a chain of several Ricci-flat subspaces. They contain, as particular cases,  the solutions previously considered by Yoshimura \cite{Yoshimura} and Myers \cite{Myers2}. Various extensions of these solutions, as well as a thorough analysis of their  singularities and horizons, are provided by Ivashchuk and Melnikov \cite{IM1}-\cite{IM3}.     
 
Although the problem of finding higher dimensional extensions of the Schwarzschild metric has been thoroughly discussed, the question of how these multidimensional solutions reduce to lower dimensions seems to have been less discussed. In this paper we study this question. In particular, in the case of ordinary $4D$ spacetime  we ask: 
(i) How does 
the physics in $4D$ depend on the number of extra dimensions?  
(ii) Does the physics in $4D$ depend on the specific nature of the singularities  of the higher-dimensional spacetime? 
(iii)  Does the physics in $4D$ depend on whether the extra dimensions are compact or large? 
(iv) Can we provide some specific observational criterion to determining  whether  the putative  extra dimensions are compact or large?   

\section{Field equations and their solutions}

 To facilitate the discussion, make the paper self-consistent and introduce our notation we start by reviewing  the solution presented in \cite{Fadeev}. In this work the spacetime signature is $(+, -, -, -)$; we follow the definitions of Landau and Lifshitz \cite{Landau}; and the speed of light $c$ is taken to be unity.

Let us write the metric as 
\begin{equation}
\label{the general metric}
dS^2 = A^2(r) dt^2 - B^2(r)\left[dr^2 + r^2 d\Omega_{(2 + n)}\right] - \sum_{i = 1}^{N}C_{(i)}^{2}(r)dg_{(i)}^2,
\end{equation}
where $d\Omega_{(n + 2)}^2$ is the metric on a unit $(n + 2)$-sphere $(n = 0, 1, 2...)$ and 
\begin{equation}
\label{definition of dg's}
dg_{(i)}^2 = \sum_{a,b = 1}^{m_{(i)}}\delta_{a b}(x_{_{(i)}})\;  dx_{(i)}^{a}dx_{(i)}^{b}, \;\;\;\; m_{(i)} \geq 1.
\end{equation}
 Here $n$ is the number of ``internal" dimensions; $N$ is the number of ``external" subspaces; $m_{(i)}$ is the dimension of the $i$-th subspace
 It is assumed that the Ricci tensors $R_{a b}(x_{(i)})$ formed out by the $\delta_{a b}(x_{(i)})$ alone all vanish.

If we introduce the quantity
\begin{equation}
\label{definition of V}
V = \prod_{i = 1}^{N}{\left[C{_{(i)}}\right]^{m_{(i)}}},
\end{equation}
then, the  field equations become
\begin{equation}
\label{R00}
\frac{A''}{A'} + (n + 1) \frac{B'}{B} + \frac{n + 2}{r} + \frac{V'}{V} = 0
\end{equation}

\begin{equation}
\label{R11}
\frac{A''}{A} + (n + 2)\frac{B''}{B} +  \frac{B'}{B}\left[\frac{n + 2}{r} - \frac{A'}{A} - (n + 2) \frac{B'}{B}\right]  - \frac{B' V'}{B V}  + \sum_{i = 1}^{N}   \frac{m_{(i)} \; C_{(i)}''}{C_{(i)}} = 0,
\end{equation}

\begin{equation} 
\label{angular part}
\frac{B''}{B} + \frac{B'}{B}\left[\frac{2n + 3}{r} + \frac{A'}{A} + n \frac{B'}{B}\right] + \frac{A'}{ r A}  + \frac{V'}{V} \left[\frac{B'}{B} + \frac{1}{r}\right]= 0,
\end{equation}

\medskip

\begin{equation}
\label{y-sector}
\frac{A'}{A} + (n + 1)\frac{B'}{B} + \frac{n + 2}{r} + \frac{V'}{V} +  \left(\frac{C_{(i)}''}{C_{(i)}'} - \frac{C_{(i)}'}{C_{(i)}}\right) = 0, 
\end{equation}
Combining (\ref{R00}) and (\ref{y-sector}) we obtain   $C_{(i)}'/C_{(i)} = - \gamma_{(i)} A'/A$, where $\gamma_{(i)}$ is a constant of integration. Thus, $C_{(i)} \propto A^{- \gamma_{(i)}}$ and   $V'/V = - \omega \; A'/A$, where $\omega = \sum_{i = 1}^N {m_{(i)}\gamma_{(i)}}$. Therefore,

\begin{equation}
\label{expression for V}
V = V_{0} A^{- \omega},
\end{equation}
where $V_{0}$ is a constant of integration. Now from (\ref{R00}) it follows that   
$B^{n + 1} \propto A^{\omega}/r^{n + 2}A'$. Substituting these expressions into (\ref{angular part}) we obtain an  equation for $A$ whose solution is

\begin{equation}
\label{A}
A(r) = \left(\frac{a r^{n + 1} - 1}{a r^{n + 1} + 1}\right)^{\alpha},
\end{equation}
where $a$ and $\alpha$ are constants of integration. Henceforth we assume $ a \neq 0$ and $\alpha \neq 0$. Consequently, the remaining metric functions are given by\footnote{The constants are chosen in such a way that the metric functions tend to unity at spatial infinity.}
\begin{equation}
\label{B, C}
B^{n + 1}(r) = \frac{\left(a r^{n + 1} + 1\right)^{[\alpha (1 - \omega) + 1]}}{a^2 r^{2 (n + 1)}\left(a r^{n + 1} - 1\right)^{[\alpha (1 - \omega) - 1]}}, \;\;\;C_{(i)}(r) = \left(\frac{a r^{n + 1} + 1}{a r^{n + 1} - 1}\right)^{\alpha \gamma_{(i)}}.
\end{equation}
Finally, in order to satisfy  (\ref{R11}) the constants of integration $\{\alpha, \gamma_{(i)}\}$ must obey the relation\footnote{To facilitate the verification of the solution we note that $\sum_{i = 1}^{N}   \frac{m_{(i)} \; C_{(i)}''}{C_{(i)}} = - \omega \left(\frac{A'}{A}\right)' + \sigma \left(\frac{A'}{A}\right)^2$}
\begin{equation}
\label{compatibility with R11}
\alpha^2 \left[\left(\omega - 1\right)^2 + \left(\sigma + 1\right) \left(n + 1\right)\right] = n + 2, \;\;\;\omega \equiv \sum_{i = 1}^{N}{m_{(i)} \gamma_{(i)}}, \;\;\;\;\sigma \equiv \sum_{i = 1}^{N}{m_{(i)} \gamma_{(i)}^2}.
\end{equation}
 In the case where $n = 0$; $N = 1$; $m = 1$, setting $\alpha = \epsilon k$ and $\gamma = 1/k$ this reduces to $\epsilon^2 (k^2 - k + 1) = 1$, which is the consistency relation in Davidson-Owen solution \cite{Davidson Owen}.

 We note that all the above quantities are invariant under the simultaneous change $a \rightarrow - a$, $\alpha \rightarrow - \alpha$. Therefore, 
 the solutions with $\alpha < 0$, $a > 0$ duplicate those with $\alpha > 0$, $a < 0$. Consequently, in what follows, without loss of generality,   we assume  $a > 0$.

The physical and geometrical properties of the solutions depend on the behavior  of  the metric functions near $a r^{n + 1} \sim 1$.  To illustrate this we first consider the physical radius $R(r)$ of a $(n + 2)$ sphere with coordinate $r$, which  is given by 
\begin{equation}
\label{physical radius}
R(r) = r B(r).
\end{equation}
 At large distances, i.e. $a r^{n + 1} \gg 1$, $R \sim r$. However, for $a r^{n + 1} \sim 1$  

\begin{eqnarray}
\label{The origin}
a R^{n + 1} \sim  2^{[\alpha (1 - \omega) + 1]}\;\; \left(a r^{n + 1} - 1\right)^{\alpha (\omega - 1) + 1}.
\end{eqnarray}
On the other hand we find

\begin{equation}
\label{gRR}
- g_{RR} dR^2 = B^2(r) dr^2 = \frac{\left[a^2 r^{2 (n + 1)} - 1\right]^2 \; d R^2}{\left[a^2 r^{2 (n + 1)} + 2 \alpha (\omega - 1) a r^{n + 1} + 1\right]^2}.
\end{equation}
 Thus, regarding the behavior near $a r^{n + 1} = 1$ there are three distinct families of solutions: (i) When $\alpha (\omega - 1) + 1 > 0$ we find that $g_{00} \rightarrow 0$, $g_{RR} \rightarrow 0$, $R(r) \rightarrow 0$ as $a r^{n + 1} \rightarrow 1^{+}$. These solutions represent naked singularities; (ii) In the same limit when $\alpha (\omega - 1) + 1 < 0$ we find  $g_{00} \rightarrow 0$, $g_{RR} \rightarrow 0 $, $R(r) \rightarrow \infty$. In these solutions  $dR/dr$ vanishes  at some finite value of $r$, say $\bar{r}$. Since $R_{min} = R(\bar{r}) > 0$, they can be used to generate  higher dimensional  wormholes  similar to those discussed in $5D$ by Agnese {\it et al} \cite{Agnese}; (iii) When   $\alpha (\omega - 1) + 1 = 0$ we find that $g_{00} \rightarrow 0$, $g_{RR} \rightarrow - \infty$, $a R^{n + 1}(r) \rightarrow 2^{[\alpha (1 - \omega) + 1]}$ as $a r^{n + 1} \rightarrow 1^{+}$. These solutions  are specially simple because now $\alpha$ is fixed. Namely,  either $\alpha = 1$, $\omega = 0$, or $\alpha = 1/(\omega - 1)$. Bellow we present  them separately.

$\bullet$ When  $\alpha = 1$, the condition\footnote{Note  that $\alpha = - 1$ is not a possible solution of  $\left[\alpha (\omega - 1) + 1\right] = 0$ because this would require $\omega = 2$, for which  (\ref{compatibility with R11}) yields $\sigma = 0$. In turn this requires $\gamma_{(i)} = 0$, i.e. $\omega = 0$ instead of $\omega = 2$.} $\alpha (\omega - 1) + 1 = 0$ yields $\omega = 0$. Then from (\ref{compatibility with R11}) it follows that $\sigma = 0$, which in turn requires $\gamma_{(i)} = 0$ for $i = (1, . . , N) $, i.e. $C_{(i)} = C_{(i)}^{0} = $ constant. In this case (\ref{physical radius}) reduces to
\begin{equation}
\label{Schwarzschild coordinate}
R = \frac{\left[a r^{n + 1} + 1\right]^{2/(n + 1)}}{a^{2/(n + 1)} r}.
\end{equation}

The physical meaning of $a$ is obtained from the asymptotic behavior of $g_{00}$. Far away from a stationary source $g_{00} \sim (1 + 2\phi)$, where $\phi$ is the Newtonian gravitational potential which goes as $- M/r^{n + 1}$, and $M$ represents the total gravitational mass. In the present case from (\ref{A}) we find 
\begin{equation}
\label{total mass}
M = \frac{2}{a}.
\end{equation}
In terms of $R$ the solution becomes 
\begin{equation}
\label{Schw solution}
dS^2 = \left(1 - \frac{2 M}{R^{n + 1}}\right)dt^2 - \left(1 - \frac{2 M}{R^{n + 1}}\right)^{- 1}dR^2 -                R^2 d\Omega_{(n + 2)}^2 - \sum_{i = 1}^{N}\left(C_{(i)}^{0}\right)^{2} dg_{(i)}^2,
\end{equation}
which, up to the innocuous $\sum_{i = 1}^{N}{m_{(i)}}$ flat extra dimensions,  describes  the so-called Schwarzschild-Tangherlini black holes  with spherical symmetry in $(n + 3)$ rather than three spatial dimensions. 
The radius $R_{h}$ of the horizon of the black hole is given by $R_{h}= (4/a)^{1/(n + 1)}$, which in isotropic coordinates corresponds to $a r_{h}^{n + 1} = 1$, as expected. For $n = 0$ they reduce to the conventional Schwarzschild solution of general relativity.

$\bullet$ For $\alpha = 1/(1 - \omega)$, from (\ref{compatibility with R11}) we find $\omega^2 - 2 \omega - \sigma = 0$. Thus, $\omega = 1 \pm \sqrt{1 + \sigma}$. The correct solution is the one that gives $\omega = 0$ when $\sigma = 0$. Consequently, $\omega = 1 - \sqrt{1 + \sigma}$ and 
\begin{equation}
M = \frac{2}{a \sqrt{1 + \sigma}}.
\end{equation} 
In terms of the Schwarzschild coordinate (\ref{Schwarzschild coordinate}) the solution becomes

\begin{equation}
\label{Tangherlini-like sol.}
dS^2 = F^{1/\sqrt{1 + \sigma}} dt^2 - \frac{dR^2}{F} - R^2 d\Omega_{(n + 2)} - \sum_{i = 1}^{N} F^{- \gamma_{(i)}/\sqrt{1 + \sigma}} dg_{(i)}^2,
\end{equation}
where 
\begin{equation}
\label{F for Tangherlini-like sol.}
F = 1 - \frac{2 \sqrt{1 + \sigma} M}{R^{n + 1}}.
\end{equation}
For $\sigma = \sum_{i = 1}^{N}m_{(i)}\gamma_{(i)}^2 = 0$, we recover  (\ref{Schw solution}).  It is important to emphasize that (\ref{Schw solution}) and (\ref{Tangherlini-like sol.})-(\ref{F for Tangherlini-like sol.}) are the {\it only} family of solutions for which $g_{tt} = 0$ and $g_{RR} = - \infty$ in the same region of the spacetime. 

 For completeness, we briefly  examine the radial motion of light towards the center. Assuming $dS = 0$  from (\ref{the general metric}) we get
\begin{equation}
\label{radial motion of light}
dt = - \left(\frac{B}{A}\right) dr. 
\end{equation}
To study the motion  near the singularity we introduce the coordinate $\xi = a r^{n + 1} - 1$ and consider an expansion about $\xi = 0$. Then (\ref{radial motion of light}) becomes
\begin{equation}
\label{radial motion of light near the singularity}
d \tau = a^{1/(n + 1)} dt \sim - \xi^{\kappa} d\xi, \;\;\;\kappa = \frac{\alpha (\omega - 2 - n) + 1}{n + 1}. 
\end{equation}
The time required to move from a point $ \xi = \xi_{0}$ in the neighborhood of the singularity to $\xi = 0$ is finite for every $\kappa \neq -1$. However, for $\kappa = -1$ we get $\xi \sim e^{- \tau}$ which means that there is a horizon. Substituting $\kappa = - 1$ into the compatibility equation (\ref{compatibility with R11}) we get $\sigma (n + 2) + \omega^2 = 0$ whose solution is $\sigma = \omega^2 = 0$ (recall that $\sigma \geq 0$). The conclusion is that only the Schwarzschild-Tangherlini spacetimes possess a horizon, all the rest are naked singularities.

\section{Dimensional reduction for compact extra dimensions}

First we study the dimensional reduction of the solutions in the case where the external  coordinates are rolled up to a small size. To put the discussion in perspective, let us recall that when  $n = 0$ the curvature scalar  $R_{(D)}$ associated 
with the metric 

\begin{equation}
\label{general metric}
dS^2 = \gamma_{\mu\nu}(x)dx^{\mu}dx^{\nu} - \sum_{i = 1}^{m}H_{i}^2(x)dy_{i}^2,\;\;\;\;m = \sum_{i = 1}^{N}{m_{(i)}},
\end{equation}
 can be expressed as (see, e.g., \cite{IM4})
\begin{equation}
\label{dimensional reduction  of the Ricci scalar}
\sqrt{|g_{D}|}R_{(D)} = \sqrt{|g_{(4)}^{\mbox{eff}}|}\; \left[R_{(4)} - \sum_{a = 1}^{m}\frac{\partial_{\mu}H_{a}\partial^{\mu} H_{a}}{N_{a}^2} - \frac{1}{2}\sum_{a = 1}^{m}\sum_{b = 1}^{m}\left(\frac{\partial_{\mu}H_{a}}{H_{a}}\right)\; \left(\frac{\partial^{\mu}H_{b}}{H_{b}}\right) + 
\Delta_{(4)}{\ln V}\right]. 
\end{equation}
Here  $R_{(4)}$ is the four-dimensional  curvature scalar calculated from the effective $4D$ metric tensor
\begin{equation}
\label{effective metric}
g_{\mu \nu}^{\mbox{eff}} = \gamma_{\mu\nu}\; \prod_{i = 1}^{m} H_{i}; 
\end{equation} 
$g_{(D)}$ and $g_{(4)}^{\mbox{eff}}$ denote the determinants of the $D$-dimensional metric (\ref{general metric}) and effective $4D$ metric (\ref{effective metric}), respectively;  $V$ is the function defined in (\ref{definition of V}), which in the notation of (\ref{general metric}) 
becomes $V =   \prod_{i = 1}^{m} H_{i}$, and $\Delta_{(4)}$ is the Laplace-Beltrami operator corresponding to $g_{\mu \nu}^{\mbox{eff}}$.  

The choice of the factor $\prod_{i = 1}^{m} H_{i}$ in (\ref{effective metric})  assures that the effective action in $4D$ contains the exact Einstein Lagrangian, with a fixed effective gravitational constant \cite{Davidson Owen}, \cite{Dolan}.  The second and third expressions in (\ref{dimensional reduction  of the Ricci scalar}) are proportional to a Lagrangian for an effective energy-momentum tensor (EMT) $T_{\mu\nu}$, while the last one  gives rise to a boundary term in the effective action and vanishes when
  the field equations $R_{AB} =0$ are imposed.
Thus, for the higher-dimensional metric (\ref{A})-(\ref{B, C}) with $n = 0$   all the physics in $4D$ is concentrated in the effective  line element $g_{\mu\nu}^{\mbox{eff}} = V \gamma_{\mu\nu}$.

Let us now go back to the case where $n > 0 $. We have found that the effective metric in $(4 + n)$-dimensions, say  $g_{{\tilde{\mu}} \tilde{\nu}}^{\mbox{eff}}$, is obtained             from $\gamma_{{\tilde{\mu}} \tilde{\nu}}$, the $(4 + n)$ part of the metric in $D$-dimensions,  as 

\begin{equation}
\label{factor for different n}
g_{{\tilde{\mu}} \tilde{\nu}}^{\mbox{eff}} = \gamma_{{\tilde{\mu}} \tilde{\nu}}\; \prod_{i = 1}^{m} H_{i}^{p(n)},\;\;\;p(n) = \frac{2}{n + 2}, 
\end{equation}
which for $n = 0$ reduces to (\ref{effective metric}). Similar to the above discussion, with this choice  the gravitational action has the standard form
\begin{equation}
\label{gravitational  action for any D}
S_{(4 + n)} = - \frac{1}{16 \pi G_{(4 + n)}}\int{\sqrt{|g_{(4 + n)}^{\mbox{eff}}|}}\; R_{(4 + n)}  d^{4 + n} x , 
\end{equation}
in any number of dimensions, where $G_{(4 +n)}$ is  the gravitational constant in $(4 + n)$. If $L_{(j)}$ represents  the size of the j-th external coordinate, then 

\begin{equation}
\label{gravitational constant in 4 + n}
\frac{1}{G_{(4 +n)}}  = \frac{\prod_{j = 1}^{m} L_{(j)}}{G_{D}}.
\end{equation}
In what follows, to simplify the notation we set $G_{(4 + n)} = 1$.

In the case under consideration $g_{\tilde{\mu}\tilde{\nu}}^{\mbox{eff}} = V^{2/(n + 2)}\gamma_{\tilde{\mu}\tilde{\nu}}$. Thus the effective gravity in $(4 + n)$ is determined by the line element

\begin{equation}
\label{effective line element for any n }
ds^2 = \left(\frac{a r^{n + 1} - 1}{a r^{n + 1} + 1}\right)^{2 \varepsilon} dt^2 - \frac{1}{\left(a r^{n + 1}\right)^{4/(n + 1)}}\frac{\left(a r^{n + 1} + 1\right)^{2 (\varepsilon + 1)/(n + 1)}}{\left(a r^{n + 1} - 1 \right)^{2(\varepsilon - 1)/(n + 1)}} \; \left[dr^2 + r^2 d\Omega^2_{(n + 2)}\right], 
\end{equation} 
where 
\begin{equation}
\label{varepsilon  for n }
\varepsilon = \frac{\alpha (n + 2 - \omega)}{n + 2}.
\end{equation}
We note that $\varepsilon^2 \leq 1$. In fact, substituting (\ref{varepsilon  for n }) into (\ref{compatibility with R11}) we find
\begin{equation}
\label{1 - varepsilon2 for any n}
1 - \varepsilon^2 = \frac{(n + 1)\left[\omega^2 + \sigma (n + 2)\right]}{(n + 2)\left[(\omega - 1)^2 + (\sigma + 1)(n + 1)\right]} \geq 0.
\end{equation}
An observer in $(4 + n)$, who is not aware of the existence of external extra dimensions, interprets the metric functions as if they were governed by an effective  energy-momentum tensor (EMT) $T_{A B}$ determined by the  Einstein field equations $G_{AB} = 8 \pi T_{A B}$. In the present case, using (\ref{effective line element for any n }) we find

\begin{eqnarray}
\label{effective EMT in isotropic coordinates, arbitrary n}
8\pi T_{0}^{0} &=& \frac{2 (1 - \varepsilon^2)(n + 1)(n + 2) \; a^{2(n + 3)/(n + 1)}\;  r^{2(2 + n)} }{\left[a^2 r^{2(n + 1)} - 1\right]^2}\left[\frac{(a r^{n + 1} - 1)^{2(\varepsilon - 1)}}{(a r^{n + 1} + 1)^{2(\varepsilon + 1)}}\right]^{1/(n + 1)}, \nonumber \\
T_{1}^{1} &=& - T_{0}^{0}, \;\;\;\;\;\;T_{2}^{2} = T_{3}^{3} = \cdots = T_{n + 3}^{n + 3} =  T_{0}^{0}.
\end{eqnarray}

By virtue of (\ref{1 - varepsilon2 for any n}),  the effective energy density $T_{0}^{0}$ results to be  automatically non-negative.
When  $\varepsilon = 1$, both $\omega$ and $\sigma$ must vanish, which in turn implies $\gamma_{(i)} = 0$. Consequently, (\ref{effective line element for any n }) reduces to Schwarzschild-Tangherlini's spacetimes in isotropic coordinates, as expected. 

The relationship between the components of the EMT suggest that the source can be interpreted as a neutral massless scalar field  
\begin{equation}
\label{The scalar field for any n}
\Psi = \int{\sqrt{- 2 g_{rr}T_{0}^{0}}\; dr}.
\end{equation}
After integration we find
\begin{equation}
\label{scalar field for any n}
\Psi(r) = \frac{1}{2}\sqrt{\frac{(n + 2)(1 - \varepsilon^2)}{2 \pi (n + 1)}}\ln \mid\frac{a r^{n + 1} - 1}{a r^{n + 1} + 1}\mid.
\end{equation}
It is not difficult to verify that $\Psi$ satisfies the Klein-Gordon equation 

\[\frac{\left(\sqrt{- g} g^{\mu\nu}\Psi_{,\mu}\right), _{\nu}}{\sqrt{- g}} = 0,\]
which 
is consistent with our interpretation. It should  be noted that (\ref{effective line element for any n }), (\ref{effective EMT in isotropic coordinates, arbitrary n}), (\ref{scalar field for any n}) for $n = 0$  are equivalent to the static, spherically symmetric solution of the coupled Einstein-massless scalar field equations originally discovered by Fisher \cite{Fisher} and rediscovered by Janis, Newman and Winicour \cite{JNW}. Thus, the above equations generalize Fisher's solution to $(4 + n)$ dimensions. 

The coordinate transformation
\begin{equation}
\label{definition of z for any n}
z = \frac{a}{2 (n + 1)} \ln\mid \frac{a r^{n + 1} - 1}{a r^{n + 1} + 1}\mid,
\end{equation}
renders the metric (\ref{effective line element for any n })  into a  form where $g_{zz}(z) = - g_{tt}(z) g_{\theta \theta}^{n + 2}(z)$. In terms of $z$ the solution of the Klein-Gordon equation is $\Psi = q z$, where $q$ is interpreted as the scalar charge \cite{BronnikovActaPhys}. Thus, from (\ref{scalar field for any n}) we find the scalar charge in the present case as 

\begin{equation}
\label{scalar charge for any n}
q = \frac{M}{2 \varepsilon}\sqrt{\frac{(n + 1)(n + 2)(1 - \varepsilon^2)}{2 \pi}}, \;\;\;M = \frac{2 \varepsilon }{a},
\end{equation}
where $M$ is the total mass measured by an observer located at spatially infinity. If we denote as $M_{ST}$ the total Schwarzschild-Tangherlini mass $(\varepsilon = 1)$, then the above implies
\begin{equation}
\label{contribution of q to M, for any n}
M^2 = M_{ST}^2 - \frac{8 \pi q^2}{(n + 1)(n + 2)}.
\end{equation}
Thus, $M \leq M_{ST}$. We note that $M \rightarrow 0$ as $\varepsilon \rightarrow 0^{+}$ and $M \rightarrow M_{ST}$ as $\varepsilon \rightarrow 1^{-}$.

The line element (\ref{effective line element for any n }) acquires a more familiar form in terms of the  radial coordinate $R$ defined by
\begin{equation}
\label{new radial coordinate R for n }
R = r\left(1 + \frac{1}{a r^{n + 1}}\right)^{2/(n + 1)}.
\end{equation}
Indeed,  it becomes

\begin{equation}
\label{effective line element for n in Schw-like coordinates} 
ds^2 = \left(1 - \frac{2 M/ \varepsilon}{R^{n + 1}}\right)^{\varepsilon} dt^2 -  \left(1 - \frac{2 M/\varepsilon}{R^{n + 1}}\right)^{- (n + \varepsilon)/(n + 1)}  dR^2 - R^2\left(1 - \frac{2 M/\varepsilon}{R^{n + 1}}\right)^{(1 - \varepsilon)/(n + 1)}d\Omega^2_{(n + 2)}. 
\end{equation} 

In addition, the effective  EMT is now given by

\begin{equation}
\label{EMT, for any n}
8 \pi T_{0}^{0} = \frac{( n + 1)(n + 2) (1 - \varepsilon^2) M^2}{2\;  \varepsilon^2 R^{2(n + 2)}}\left(1 - \frac{2 M/\varepsilon}{R^{n + 1}}\right)^{(\varepsilon - n - 2)/(n + 1)}, \;\;\;T_{1}^{1} = - T_{0}^{0}, \;\;\;\;\;\;T_{2}^{2} = T_{3}^{3} = \cdots = T_{n + 3}^{n + 3} =  T_{0}^{0}. 
\end{equation}

It should be noted that the dimensional reduction {\it eradicates} the geometrical and physical differences between the three   families of higher-dimensional  solutions derived  in (\ref{The origin})-(\ref{gRR}). Specifically, the effective $(4 + n)$ spacetime shows no evidence of the different  nature of the singularity of $g_{RR}$ near $a r^{n + 1} = 1$ in higher dimensions. The fact is that all solutions in (\ref{A})-(\ref{B, C}) generate the same effective spacetime in $(4 + n)$, regardless of their specific properties.   

 From (\ref{EMT, for any n}) it follows that   $T = ( n+ 2) T_{0}^0$. As a consequence all the components of the Ricci tensor, except for $R_{11}$,  are zero\footnote{From the field equations we get $R_{A B} = 16 \pi \delta_{A}^{1} T_{1 B}$.}.  
We now recall that in the case of a constant, asymptotically flat,  gravitational field there is an expression for the total energy of matter plus field, which is an integral of $R_{0}^{0}$ over the volume ${\cal{V}}$ occupied by the matter\footnote{In Landau and Lifshitz \cite{Landau} the discussion is in $4D$, however it can be extended to any number of dimensions} \cite{Landau}, viz.,  $M = \kappa \int{\sqrt{|g|}R_{0}^{0}d{\cal{V}}}$, where the constant of proportionality $\kappa$ depends on the number of dimensions, e.g., $\kappa = 1/4\pi$ in $4D$. In conventional general relativity this expression  is known as the Tolman-Wittaker formula.
Since $R_{0}^{0} = 0$, it follows that the gravitational mass of any spherical shell is just zero.  This conclusion holds for any $n$ and $\varepsilon$, including the Schwarzschild-Tangherlini black holes, as well as the familiar Schwarzschild solution of general relativity. 

We note that $R_{0}^{0} = 0$ implies
\begin{equation}
\label{nongravitating matter}
(n + 1) \rho + p_{r} + (n + 2)p_{\perp} = 0, \;\;\; p_{r} = - T_{1}^{1}, \;\;\;p_{\perp} = - T_{2}^{2},
\end{equation}
which generalizes to $n$ dimensions the well-known equation of state $\left(\rho + p_{r} + 2 p_{\perp}\right) = 0$  for nongravitating matter in $4D$, which in turn generalizes to anisotropic matter the  equation of state $(\rho + 3 p) = 0$ for a perfect fluid that has no effect on gravitational interactions \cite{WessonEssay}-\cite{Presentauthor}.

The case where $n = 0$ is especially important because it corresponds to our spacetime with spherical symmetry in the three usual spatial dimensions, 

\begin{equation}
\label{effective 4D solution in a new parameterization}
ds^2 = \left(1 - \frac{2 M/\varepsilon}{ R}\right)^{\varepsilon}\; dt^2 - \frac{dR^2}{\left(1 - \frac{2 M/\varepsilon}{ R}\right)^{\varepsilon}} - R^2 \left(1 - \frac{2 M/\varepsilon}{ R}\right)^{1 - \varepsilon}\left(d\theta^2 + \sin^2\theta d\phi^2\right).
\end{equation}
It has two distinctive properties, which {\it do not} hold for any other $n \neq 0$: (i) The effective spacetime is Schwarzschild-like in the sense that the $4D$  line element  is in the ``gauge" $g_{00}\; g_{11} = -1$, which has a number of important properties and applications \cite{Bronnikov1}-\cite{Bronnikov3};  (ii) At large distances from the origin, to first order in $M/R$,  it is indistinguishable from the Schwarzschild vacuum exterior for any $\varepsilon$. As a consequence,  (\ref{effective 4D solution in a new parameterization}) is compatible with   the (weak) equivalence principle,  for {\it all} values of $\varepsilon$. In addition, in the weak-field approximation, it is  consistent with   Newtonian physics \cite{JPdeLCQG}, \cite{JPdeLIJMPD}.  

Horizons and singularities in static spherically symmetric configurations in $4D$  were recently discussed and classified by Bronnikov {\it et al} \cite{Bronnikov2}, \cite{Bronnikov3} in terms of the quantity 

\begin{equation}
\label{Z tilde}
\tilde{Z} \propto \frac{(R^{12}_{\;\;\;\;12} - R^{02}_{\;\;\;\;02})}{g_{00}},
\end{equation} 
which characterizes the magnitude of tidal forces in a freely falling reference system near the spacetime region where $g_{tt} = 0$. In the present case it yields  
\begin{equation}
\label{Z tilde}
\tilde{Z} \propto \frac{(1 - \varepsilon^2) M^2}{\varepsilon^2 R^4 \left(1 - 2 M/\epsilon R\right)^2}
\end{equation} 
Thus,  $\tilde{Z} = 0$ for $\varepsilon = 1$ (the Schwarzschild black hole) and $\tilde{Z} \rightarrow \infty$, as $g_{00} \rightarrow 0$,  for any other $\varepsilon$. Consequently, in the classification given in \cite{Bronnikov2} for $\varepsilon \neq 1$ the singularity at $R = 2 M/\varepsilon$ is a ``truly naked" one.

\section{Splitting procedure for large extra dimensions}

We now assume that the external dimensions are large. In order to  make contact with previous works,  we first consider  $N = 1$, $m = 1$. To perform the splitting of the spacetime into $(4 + n) + 1$ we introduce a unit vector $\chi^{A}$ tangent to the extra dimension, and assume that the $(4 + n)$ manifold is locally orthogonal to the large extra dimension. As a consequence the metric induced on $(4 + n)$ is given by $g_{C D} = \gamma_{C D} - \chi_{C}\chi_{D}$, which in the present case is just the $(4 + n )$ part of (\ref{the general metric}) with $A(r)$ and $B(r)$ given by (\ref{A}) and (\ref{B, C}), respectively, viz., 

\begin{equation}
\label{metric in 4+n, isotropic coordinates, one large extra dimension}
ds^2 = \left(\frac{a r^{n + 1} - 1}{a r^{n + 1} + 1}\right)^{2 \alpha} dt^2 - \frac{1}{\left(a r^{n + 1}\right)^{4/(n + 1)}}\frac{\left(a r^{n + 1} + 1\right)^{2 [\alpha(1 - \omega) + 1]/(n + 1)}}{\left(a r^{n + 1} - 1 \right)^{2[\alpha(1 - \omega) - 1]/(n + 1)}} \; \left[dr^2 + r^2 d\Omega^2_{(n + 2)}\right]
\end{equation} 
 
Following a splitting procedure similar to the one used in \cite{WessonPoncedeLeon1}  we obtain an effective EMT in $(4 + n)$

\begin{equation}
8\pi T_{A B} = \frac{C_{A;B}}{C},
\end{equation}
where $C_{A} = \partial C/\partial x^{A}$. In the present case ( $N = 1$, $m = 1$) $\omega = \gamma$, $\sigma = \gamma^2$. Therefore $C$ satisfies the Klein-Gordon equation constructed with the metric (\ref{metric in 4+n, isotropic coordinates, one large extra dimension}). Consequently, the trace of the EMT vanishes and   
 the Ricci scalar is zero. The nonvanishing components of the EMT are\footnote{It is worth mentioning  that the effective matter quantities do not have to satisfy the regular energy conditions because they involve terms of geometric origin \cite{Bronnikov4}.}

\begin{eqnarray}
\label{effective EMT in isotropic coordinates, arbitrary n. Second interpretation}
8\pi T_{0}^{0} &=&   \frac{2 (n + 1)(n + 2) \; \;\left[1 - \alpha^2 (\gamma - 1)^2\right]\;\;  a^{2(n + 3)/(n + 1)}\;  r^{2(2 + n)}  }{\left[a^2 r^{2(n + 1)} - 1\right]^2\;\;}\left[\frac{(a r^{n + 1} - 1)^{2[\alpha (1 - \gamma) - 1]}}{(a r^{n + 1} + 1)^{2[\alpha (1 - \gamma) + 1]}}\right]^{1/(n + 1)},\nonumber \\ 
T_{1}^{1} &=& - T_{0}^{0}\;\;\left\{1 + \frac{\alpha \gamma \left[a^2 r^{2(n + 1)} + 1\right]}{a r^{n + 1}\left[1 - \alpha^2 (\gamma - 1)^2\right]}\right\}, \nonumber \\
T_{2}^{2} &=& - \frac{T_{0}^{0} + T_{1}^{1}}{n + 2},
\end{eqnarray} 
From the compatibility condition (\ref{compatibility with R11})  we find 
\begin{equation}
1 - \alpha^2(\gamma - 1)^2 = \frac{2\gamma \alpha^2 (n + 1)}{(n + 2)}.
\end{equation}
Thus, (i) If  $\gamma = 0$, then $\alpha^2 = 1$ and the EMT vanishes. Consequently, the metric reduces to the Schwarzschild-Tangherlini spacetime with mass $M_{ST} = 2/a$; (ii) If $\gamma > 0$, then  $T_{0}^{0} > 0$ and $[1 - \alpha(1 - \gamma)] > 0$ for any $n$. As a consequence $g_{22} \rightarrow 0$ as $R^{n + 1} \rightarrow 2 M/\alpha $; (ii) If $\gamma < 0$, then $T_{0}^{0} < 0$ and $[1 - \alpha(1 - \gamma)] < 0$. Thus the metric has wormhole-like structure in the sense that $g_{22} \rightarrow - \infty$ as $R^{n + 1} \rightarrow 2 M/\alpha $, similar to those discussed for $n = 0$ in \cite{Agnese}; (iii) For $0 < \gamma < 2/(n + 2)$, we find $\alpha > 1$. In this range
\begin{equation}
\label{M for alpha > 1}
M_{ST} \leq M \leq \frac{M_{ST} (n + 2)}{\sqrt{n^2 + 4n + 3}}.
\end{equation}

In  five-dimensional  Kaluza-Klein theory $(n = 0, N = 1, m = 1)$, the line element (\ref{metric in 4+n, isotropic coordinates, one large extra dimension}) plays a central role in the discussion of  
many important observational problems, which include the classical tests of relativity, as well as the geodesic precession of a gyroscope and possible departures from the equivalence principle \cite{WessonBook}-\cite{HongyaOverduin}. In the context of the induced-matter approach, the configuration of matter (\ref{effective EMT in isotropic coordinates, arbitrary n. Second interpretation}) is interpreted as describing extended spherical objects called solitons \cite{WJPdeL} (for a recent discussion see Ref. \cite{JPdeL} and references therein). Solitons  and black holes are alike in one important aspect: they contain  a curvature singularity at the center of ordinary space.  However, (1) solitons do not have an event horizon; and (2) they have an extended matter distribution rather than having all their matter compressed into the central singularity \cite{OW}.

\medskip 

In general, for any $N$ and $m_{(i)}$ we can proceed in a similar way. However, now the trace of the effective EMT does not longer vanish, except in the case where $\omega^2 = \sigma$. 
The metric and the  effective EMT in terms of the radial coordinate $R$ introduced in (\ref{new radial coordinate R for n }) are given by

\begin{equation}
\label{metric in 4 + n, in terms of R, one large extra dimension}
ds^2 = \left(1 - \frac{2 M/\alpha}{R^{n + 1}}\right)^{\alpha}\; dt^2 - \left(1 - \frac{2 M/\alpha}{R^{n+1}}\right)^{- [n + \alpha (1 - \omega)]/(1 + n)} dR^2 -  R^2 \left(1 - \frac{2 M/\alpha}{R^{n + 1}}\right)^{[1 - \alpha(1 - \omega)]/(n + 1)}\; d\Omega_{(n + 2)}, 
\end{equation}
where $M = 2 \alpha/a$;
and
\begin{eqnarray}
8 \pi T_{0}^{0} &=& \frac{( n + 1)(n + 2) \left[1 - \alpha^2 (1 - \omega)^2\right] M^2}{2\;  \alpha^2 R^{2(n + 2)}}\left(1 - \frac{2 M/\alpha}{R^{n + 1}}\right)^{[\alpha (1 - \omega) - n - 2]/(n + 1)}, \nonumber \\
T_{1}^{1} &=& - T_{0}^{0} \; \frac{1 + (\omega^2 - 1)\alpha^2 + \omega \alpha (a R^{n + 1} - 2)}{1 - \alpha^2 (\omega - 1)^2}, \nonumber \\
T_{2}^{2} &=& - T_{0}^{0} \; \frac{[n (1 + \omega^2) + 2]\alpha^2 - \omega \alpha(a R^{n + 1} - 2) - (n + 2)}{(n + 2)[1 - \alpha^2 (\omega - 1)^2]}.
\end{eqnarray}
The effective energy density is positive in the range $(\alpha - 1)/\alpha < \omega < (\alpha + 1)/\alpha$. Wormhole-like solutions are obtained in the region $\omega < (\alpha - 1)/\alpha$.

The trace of the EMT is 
\begin{eqnarray}
T = T_{0}^{0} + T_{1}^{1} + (n + 2)T_{2}^{2} &=&  T_{0}^{0} \; \frac{(n + 2)(\sigma - \omega^2)}{2\omega + \sigma - \omega^2},\nonumber \\
&=& \frac{(\sigma - \omega^2)(n + 1)^2 (n + 2) M^2}{16 \pi R^{2 (n + 2)}}\left(1 - \frac{2 M/\alpha}{R^{n + 1}}\right)^{[\alpha (1 - \omega) - n - 2]/(n + 1)},
\end{eqnarray}
where we have used (\ref{compatibility with R11}) to eliminate $\alpha$. Thus, the effective matter quantities satisfy the equation of state 

\begin{equation}
\label{equation of state for large extra dimensions}
\rho = \frac{2 \omega + \sigma - \omega^2}{2 \omega - (n + 1)(\sigma - \omega^2)}\; \left[p_{r} + \left(n + 2\right)\; p_{\perp}\right]
\end{equation}
As it was mentioned above $T = 0$ only if $\sigma = \omega^2$. Besides, when $\omega = 0$, but $\sigma \neq 0$, the above reduces to the massless scalar field given by (\ref{effective line element for n in Schw-like coordinates} )-(\ref{EMT, for any n}), in agreement  with the fact that the factor $V$ given by (\ref{expression for V}) is constant in this case. When $\omega = 0$ and $\alpha = 1$ from (\ref{compatibility with R11})
 we find $\sigma = 0$, i.e. $\gamma_{(i)} = 0$. Thus we recover Schwarzchild-Tangherlini spacetimes.

Observations suggest   that Kaluza-Klein corrections to general relativity should be small. This means that in practice we should expect $\sigma \approx \omega^2$; $\sigma \approx 0$. Therefore we can expand (\ref{compatibility with R11}) about $\omega = 0$,
\begin{equation}
\label{series for alpha}
\alpha = 1 + \frac{\omega}{n + 2} + O(\omega^2).
\end{equation}
In this approximation $T_{0}^{0} > 0$ requires $\omega > 0$. Since $M = \alpha M_{ST}$, it follows that for positive effective density $M > M_{ST}$ (for wormhole-like distributions $M <  M_{ST}$). 

\section{Summary} 
At this point we see that the answers to the questions posed in the introduction crucially depend on whether the external extra dimensions are compact or large. For compact external extra dimensions: (i) The physics in $4D$, which is governed by  the metric (\ref{effective 4D solution in a new parameterization}),  is independent of $N$ and $m$; (ii) Since $0 < \varepsilon < 1$, the reduction procedure flattens out the rich diversity of higher dimensional solutions. In particular, the energy density is always positive and $g_ {22} = 0$ at  $ a r = 1$. 
For large extra dimensions, the situation is much more  elaborated. Firstly, from (\ref{equation of state for large extra dimensions}) we see that the effective matter quantities satisfy an equation of state that explicitly depends on $\omega$ and $\sigma$, i.e.  the number of extra dimensions. Secondly, 
the effective spacetime in $4D$ does inherit the singularities of its higher-dimensional counterpart. However, in the exceptional case where  $\omega = \sum_{i = 1}^{N}{m_{(i)}\gamma_{(i)}} = 0$ both compact and large extra (external) dimensions yield the same physics in $4D$.  

An important result here is that the total gravitational mass $M$ is different in both cases. Namely, it is  less than the Schwarzshild mass for compact extra dimensions, but bigger than the Schwarzshild mass for large extra dimensions (with the exception of wormhole-like distributions). This result may be relevant to  observations for an experimental/observational test of the theory.

\end{document}